\documentclass[sigconf,nonacm]{acmart}
\AtBeginDocument{%
  }

\copyrightyear{2026}
\acmYear{2026}
\setcopyright{cc}
\setcctype{by}
\acmConference[FTA '26]{Proceedings of the 1st International Workshop on Firmware Testing and Analysis}{October 04--09, 2026}{Oakland, CA, USA}
\acmBooktitle{Proceedings of the 1st International Workshop on Firmware Testing and Analysis (FTA '26), October 04--09, 2026, Oakland, CA, USA}
\acmDOI{10.1145/3842651.3843186}
\acmISBN{979-8-4007-2969-0/2026/10}

\usepackage{amsmath,amsfonts}
\usepackage{algorithmic}
\usepackage{booktabs}
\usepackage{multirow}
\usepackage{booktabs}
\usepackage{graphicx}
\usepackage{threeparttable}
\usepackage{graphicx}  
\usepackage{graphicx}
\usepackage{url}
\usepackage{hyperref}
\usepackage{placeins}
\usepackage{textcomp}
\usepackage{orcidlink}
\usepackage{xcolor}
\usepackage{tcolorbox}
\tcbuselibrary{skins, breakable}
\usepackage{enumitem}
\usepackage{amsmath}
\usepackage{tcolorbox}
\tcbuselibrary{skins, breakable}
\usepackage[T1]{fontenc}
\usepackage[utf8]{inputenc}
\usepackage{xcolor}
\usepackage{tcolorbox}

\definecolor{RoyalBlueBox}{RGB}{0,51,153}
\begin{document}

\title{An Empirical Study of the TianoCore Community}
\author{Nazanin Siavash}
\email{nsiavash@uccs.edu}
\orcid{0009-0000-4177-0632}
\affiliation{
  \institution{University of Colorado Colorado Springs (UCCS)}
  \state{Colorado}
  \country{United States}
}
\author{Connor Glosner}
\email{cglosne@purdue.edu}
\orcid{0009-0009-6314-331X}
\author{Ayushi Sharma}
\email{sharm616@purdue.edu}
\orcid{0009-0008-8967-8016}

\affiliation{
  \institution{Purdue University}
  \state{Indiana}
  \country{United States}
}

\author{Bianca Trinkenreich}
\email{bianca.trinkenreich@colostate.edu}
\orcid{0000-0001-7302-6082}
\affiliation{
  \institution{Colorado State University}
  \state{Colorado}
  \country{United States}
}

\author{Terrance E. Boult}
\email{tboult@uccs.edu}
\orcid{0000-0001-5007-2529}
\affiliation{
  \institution{University of Colorado Colorado Springs (UCCS)}
  \state{Colorado}
  \country{United States}
}

\author{Aravind Machiry}
\email{amachiry@purdue.edu}
\orcid{0000-0001-5124-6818}
\affiliation{
  \institution{Purdue University}
  \state{Indiana}
  \country{United States}
}
\author{Armin Moin}
\email{moin@purdue.edu}
\orcid{0000-0002-8484-7836}
\affiliation{
  \institution{Purdue University}
  \state{Indiana}
  \country{United States}
}

\renewcommand{\shortauthors}{Siavash et al.}

\begin{abstract}
We investigate the software security and maintenance practices adopted by stakeholders in the TianoCore community and identify opportunities to improve firmware development workflows. We conduct a survey and a limited interview study with participants representing independent firmware vendors, original equipment manufacturers, security experts, firmware developers, and academic researchers. This open-source development community maintains a reference implementation for the core of the UEFI firmware. We highlight important gaps in the current state of firmware development within the TianoCore ecosystem and identify key areas in which improved security practices, greater adoption of memory-safe technologies, and increased automation of manual processes could strengthen the maintenance and security of the UEFI firmware.
\end{abstract}




\maketitle

\section{Introduction}\label{sec:Introduction}
The TianoCore community maintains the open-source core of a critical piece of code, the Unified Extensible Firmware Interface (UEFI) firmware, which is widely used across numerous devices worldwide. We conduct an empirical study of security engineering practices and software maintenance workflows within the TianoCore ecosystem. We adopt a survey-based methodology~\cite{acmSurvey}, complemented by follow-up interviews. The study investigates various dimensions, including bug management, patch management, as well as verification and validation. This study is guided by the following Research Questions (RQs): \textbf{RQ1.} How do the companies and organizations in the TianoCore ecosystem structure their bug reporting and issue-tracking workflows, and what access models govern contributor and public interaction with these systems? \textbf{RQ2.} What criteria and processes do TianoCore practitioners use to triage (e.g., prioritize) bugs, and to what extent have automated or AI-assisted approaches been adopted for that? \textbf{RQ3.} Under what conditions do the companies and organizations in the TianoCore ecosystem propagate patches to upstream and downstream repositories, and what mechanisms and tooling govern cross-repository patch management? \textbf{RQ4.} What static analysis, dynamic analysis, and formal verification methods and tools are employed in the TianoCore firmware development, and what are the primary technical and organizational barriers preventing broader adoption of such methods and tools? \textbf{RQ5.} How do the companies and organizations in the TianoCore ecosystem prioritize firmware security threat categories and allocate resources to security activities, and how does TianoCore’s security posture compare to alternative firmware implementations? \textbf{RQ6.} What are the primary technical and organizational barriers to adopting memory-safe variants, particularly using Rust, in EDK II-based projects, and what level of practitioner interest and readiness currently exists concerning that? \textbf{RQ7.} To what extent have DevSecOps practices been adopted in the TianoCore/UEFI firmware ecosystem, and what aspects of bug management have been incorporated into the CI/CD pipelines? \textbf{RQ8.} How do the companies and organizations in the TianoCore/UEFI ecosystem verify the integrity of upstream components and third-party dependencies, and what are the perceived commercial and educational impacts of firmware vulnerabilities on their operations?


\section{Related Work}\label{sec:related-work}
Ayala et al. \cite{Ayala+2025} conducted a mixed method study that included surveys and interviews with maintainers of some open-source projects. Their study identified several challenges affecting OSS security engineering efforts. They also observed that many maintainers relied heavily on manual maintenance processes, despite the availability of automated tools and features. Furthermore, Alomar et al. \cite{Alomar+2020} presented an empirical investigation of vulnerability management practices based on interviews with security professionals. Finally, Wermke et al. \cite{Wermke+2022} conducted an empirical interview study with contributors, maintainers, team leaders, and project owners from a diverse set of OSS projects to investigate security and trust practices in OSS development. 

\section{Survey Results}\label{sec:survey-results}
We had 13 participants in our self-administered, anonymous, online survey. They were from the TianoCore/UEFI ecosystem and 2 of them volunteered to participate in follow-up interviews with deeper discussions. Our survey questionnaire and data are published at \cite{anonymous2026dataset}. The results suggest that firmware security is treated as a high-priority concern, although many security and maintenance workflows remain highly manual and inconsistently standardized across organizations. With respect to RQ1, bug reporting and issue-tracking practices are moderately structured but heterogeneous across the ecosystem. GitHub Issues emerged as the most commonly used issue tracking platform, although no single issue tracker dominates across all organizations. Access to the issue trackers of companies is also frequently restricted. Regarding RQ2, the respondents consistently ranked security vulnerabilities as the highest-priority bug category, demonstrating strong security awareness within the ecosystem. Bug triage remains predominantly manual, with very limited adoption of AI-assisted workflows. The software maintenance and security process maturity appears to correlate more strongly with organizational scale than with individual experience, as larger teams generally reported more structured workflows and greater tooling adoption. The findings related to RQ3 indicate that patch propagation and upstream maintenance remain largely manual processes. Security bugs are the primary trigger for upstream patch submission, and most teams prefer maintaining close synchronization with upstream repositories rather than selectively cherry-picking patches. Semantic patching and automated patch propagation techniques are not yet adopted, suggesting that cross-repository maintenance and security continues to rely heavily on developer expertise and manual coordination. With respect to RQ4, dynamic analysis approaches, such as fuzzing and system testing, are well established within firmware development workflows, whereas formal verification and advanced static analysis tooling remain limited. Despite moderate awareness of tools, such as CodeQL and HBFA, no respondent reported active adoption of them. We know from our interactions with the ecosystem actors that there is a basic use of CodeQL in some organizations but they may have not been represented in the sample. Integration of static analysis into CI pipelines was observed primarily among the most experienced respondents. The findings addressing RQ5 show that bootloader vulnerabilities are consistently perceived as the highest-priority firmware security threat. Larger organizations also report substantial investment in security-related Software Engineering efforts. Nevertheless, no respondent reported operating a formal bug bounty program, indicating that external vulnerability discovery mechanisms remain underdeveloped within this firmware ecosystem. Regarding RQ6, the respondents expressed considerable interest in memory-safe firmware development approaches, particularly Rust-based alternatives for EDK~II. However, practical adoption remains limited due to concerns related to toolchain maturity, integration with legacy C codebases, and a lack of in-house expertise. These findings suggest that interest in memory safety currently exceeds the ecosystem's technical readiness for widespread deployment. The results associated with RQ7 indicate that DevSecOps adoption remains limited overall and appears more common among less experienced practitioners than among senior respondents. This pattern may reflect generational differences in education and training rather than deliberate organizational transformation efforts. Importantly, security expertise alone does not appear to predict CI/CD or DevSecOps adoption. Finally, the findings related to RQ8 reveal that supply-chain verification practices remain predominantly manual, with respondents relying primarily on code inspection, vulnerability feed monitoring, and internally maintained dependency documentation. Dependency complexity, limited tooling support, and the absence of standardized metadata were identified as the primary barriers to effective supply-chain security. 

\section{Conclusion \& Future Work}\label{sec:conclusion-future-work}
Our findings suggested that while the TianoCore community demonstrated strong awareness of security concerns, numerous opportunities exist to improve automation, verification and validation infrastructure, vulnerability management processes, DevSecOps incorporation, and support for memory-safe development. Future studies may explore other aspects or investigate how the ecosystem may decide to address the issues pointed out here. Finally, one of the limitations of this study roots in the relatively small sample. 


\begin{acks}
This material is based upon work supported by the U.S. National Science Foundation (NSF) under Grant No. 2534021. Any opinions, findings, conclusions, or recommendations expressed in this material are those of the authors and do not necessarily reflect the views of the NSF. This work is accepted to be presented in the FTA 2026 workshop but is not published in the proceedings, according to the ACM SIGSOFT policy (https://www2.sigsoft.org/policies/pcpolicy/) that does not allow the work of organizers to be published in the workshop proceedings.
\end{acks}

\bibliographystyle{ACM-Reference-Format}
\bibliography{refs}
\end{document}